\documentclass[preprint,12pt]{elsarticle}

\usepackage{amssymb}

\journal{Journal Of Computational Physics}

\begin{document}

\begin{frontmatter}



\title{A Report of a Significant Error On a Frequently Used Pseudo Random Number Generator}


\author{Ay\c{s}e Ferhan Ye\c{s}il}
\author{M.~Cemal Yalabik}
\address{Department of Physics, Bilkent University, 06800
Ankara, Turkey}
\date{\today}

\begin{abstract}
Emergence of stochastic simulations as an extensively used computational tool for scientific purposes intensified the need for more accurate ways of generating sufficiently long sequences of uncorrelated random numbers. Even though several different 
methods have been proposed for this end, deterministic algorithms known as pseudo-random number generators (PRNGs) emerged to be the most widely used tool as a replicable, portable and easy to use method to generate such random number sequences. 
Here, we introduce a simple Poisson process whose simulation gives systematic errors when the very commonly used random number generator of the GNU C Library (Glibc) is utilised. The PRNG of Glibc is an additive lagged Fibonacci generator, 
the family of such PRNGs are accepted as relatively safe among other PRNGs. The systematic errors indicate complex correlation relations among random numbers which requires a further explanation.

\end{abstract}

\begin{keyword}
pseudo-random numbers, pseudo-random number generator,  
\PACS{02.70.-c, 05.10.-a}

\end{keyword}

\end{frontmatter}
\section{Introduction}
Random numbers are at the core of stochastic simulations for scientific purposes. Nearly in all of these cases, but especially in Monte Carlo simulations, long sequences of uncorrelated, uniformly distributed random numbers are required. 
Several different methods are used to generate such random number sequences, but as a replicable, portable and easy to use method, deterministic algorithms, namely pseudo-random number generators (PRNGs), are preferred extensively. 
However, to be confident of such a number sequence being ``sufficiently" random, {\em i.e.} any combination of random numbers within the sequence is uncorrelated, is not straightforward since there is no complete and universal statistical 
test of analysis. In other words, even though one cannot be confident that a sequence is totally random by any analysis known so far, it can still be decided that the sequence is prone to correlations in the case of empirical or theoretical 
tests indicating in that direction \cite{Thesis}.  As a result, correlation tests for PRNGs have become necessary tools for physicists \cite{Ferrenberg,Syracuse,Grassberger, schmid,Ziff,vatt,entropy,Katz,Knuth,vatt2,ran3}. Exemplifying the 
significance of these tests, it is important to note that most of the tests are direct results of PRNGs that go wrong. One of the first such paradigmatic examples, dated back to 1993, is encountered on the simulation of 2-D Ising Model which 
shows that results of various algorithms are different from the exact result and they also differ from algorithm to algorithm depending on the PRNGs used. This analysis showed that Generalized Feedback Shift Register (GFSR) PRNGs (a type of 
lagged Fibonacci generators (LFGs) which utilise exclusive-or operations) with smaller lags produce dramatically wrong results, up to hundred times of the standard deviation, when used with Wolff algorithm  \cite{Ferrenberg}. The depth first 
nature of the algorithms are suspected to interfere with the intrinsic triple correlations of the GFSRs. However, Schmid {\em et al} showed that even though the algorithm is randomly updating the system one spin at a time, if the model has triple 
structure in itself such as Blume-Capel model, use of random numbers produced by GFSR in the simulations lead to dramatic errors \cite{schmid}. On the other hand, GFSRs are not the only faulty PRNGs. LFGs with addition and subtraction operations 
also cause systematic errors if they have small lags \cite{Syracuse}. The default random number generator of GlibC, the library of the commonly used C-compiler unix based environments, is an LFG with addition operation LFG(31,3,+). It's lag is 31
which is generally considered as small.  
Grassberger reported of correlations in Lagged Fibonacci Generators with correlation lengths proportional to their lags. His method resembles Vattulainen's n-block method\cite{vatt2}. Grassberger proposed that it might be related to the 
triple correlations intrinsic to the random number sequence produced by any LFG. He also showed while simulating 3-D self-avoiding walks on a cubic lattice, that even an LFG with lag 250, LFG(250,103,+), gives poor results. This led to the 
suspicion that LFG(55,24,+) and LFG(17,5,+) may operate poorly, and was demonstrated to be so \cite{Grassberger}. In light of these, the default PRNG of GlibC should be considered as unreliable since it has a small lag as well. 

In this paper, a simple test model, which shows strong correlations when utilised with the default random number generator provided with the C compiler of Linux distributions, is introduced. C is a very commonly used programming language
and is also very popular among scientific programming communities. The GNU C Library (Glibc) is used as the C library in the GNU systems and most systems with the Linux kernel. The PRNG of this library is constructed using a linear additive 
feedback method. We demonstrate the existence of relatively strong correlations through the use of a simple algorithm which is related to a one dimensional diffusive gas problem. 
 
To inform the unspecialised reader we provide here further information about the pseudo-random number generators: There are basically two common types of PRNGs: Linear Congruential Generators (LCGs) and Lagged Fibonacci 
Generators (LFGs) \cite{Syracuse}.
LCGs are associated with the general equation
\begin{equation}
X_i = (A X_{i-1} + B) \bmod{M}
\end{equation} 
which can be represented in short as $LCG(A,B,M)$, with a maximum period of $M$ that can be reached through a suitable choice of $A$ and $B$. Here $A$,  $B$ and $M$ are integers, and  $X_i$ is the 
$i$th  generated random number. \\
LFGs have the general equation:
\begin{equation}
X_i = X_{i-P} \odot X_{i-Q}
\end{equation} 
where $P$ and $Q$ are the lags, conventionally chosen as $P > Q$. LFGs can be represented as $LFG(P,Q,\odot)$ \cite{Syracuse}. The symbol $\odot$ represents any binary arithmetic operation such as addition, subtraction, multiplication or 
exclusive-or (xor) operation. If this operation is addition or subtraction, the maximum period that can be reached for $b$-bit precision is $(2^P-1)2^{b-1}$. If the arithmetic operation is multiplication, the maximum period 
is $(2^P-1)2^{b-3}$ \cite{Syracuse}. Moreover, if the operator is xor the maximum period becomes $2^{P-1}P$. 
The random number generator of Glibc uses an additive lagged Fibonacci generator, which is a combination of a LCG and LFG utilizing 32-bit integer operations. The lags are  produced by a linear congruential generator $LCG(16807,0,2^{31}-1)$ and 
then by using those lags a $LFG(31,3,+)$ produces the random number sequence. (The first $344$ members of the sequence are omitted.)  
Newest versions of Glibc allow initialization of lags to be done by any seed ($X_0$), but here we stick to the default case in which the initial seed is equal to $1$.
In order to apply Eqn 2., an initial set of $X_i$ for $i < P$ must be defined. This is accomplished by
\begin{eqnarray}
 X_0 &=& 1 \nonumber \\
 X_i &=& 16807X_{i-1} \bmod{(2^{31}-1)} ~\mbox{for} ~ 1\le i < 31 \nonumber \\
 X_i &=& X_{i-31}  ~\mbox{for} ~i = 31,...,33 .
\end{eqnarray}
The generation part is,
\begin{equation}
 X_i = ( X_{i-3} + X_{i-31} ) \bmod{2^{32}}  ~~\mbox{for} ~~i \ge 34 . 
\end{equation}

All $X_i$'s for $0 \le i \le 343 $ are discarded and in the end $i^{\mbox{th}}$ output of the RNG becomes $R_i = X_{i+344} \div 2 $ \cite{Peter}. The integer division by $2$ is carried out to convert the number to a $31$ bit number, 
a leading zero bit is necessary to obtain a positive number.  In order to obtain real pseudo random numbers, the result is divided by the biggest $31$ bit integer: 
\begin{equation}
 r_i = \frac{R_i}{2^{31}-1}.
\end{equation}
This result then provides a pseudo-random real number linearly distributed between $0$ and $1$ with $31$ bit precession.\\

\section{Results}
\begin{figure}[!h]
\begin{center}
\includegraphics[scale=0.35]{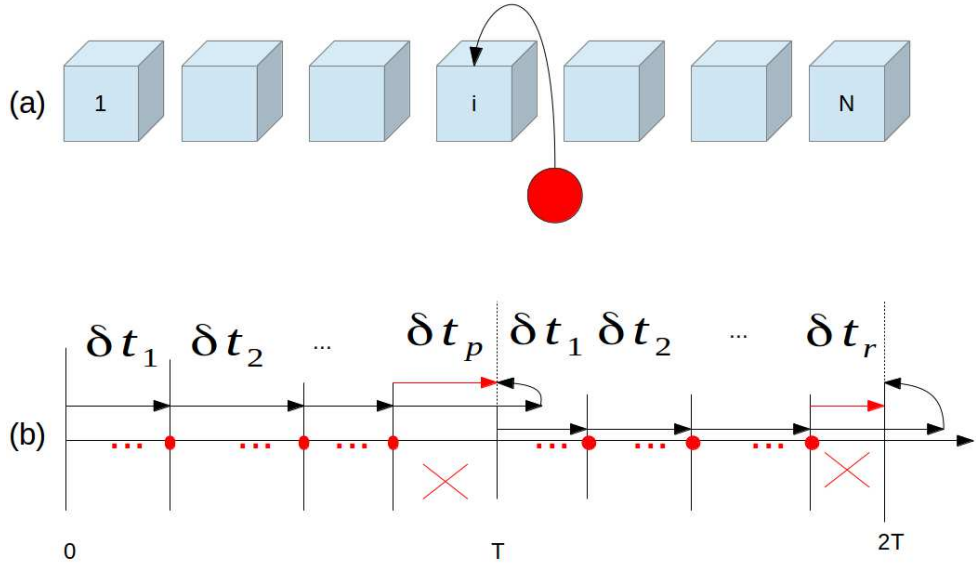}
\caption{\small a) Each box is equally likely to receive a ball at a random time. b) Time $t$ is incremented by $\delta{t}$ at each step. The small dots on the axis indicate n draws in vain from the PRNG, and the larger dots indicate a ball
is placed to a random box at that point in time. Whenever $\delta{t}$ crosses a multiple of $T$, the process is paused, i.e., no ball is placed to any box and time is set to that multiple of $T$. \protect\label{fig:time_scheme}}
\end{center}
\end{figure}

\begin{figure}[!h]
\begin{center}
\includegraphics[scale=0.85]{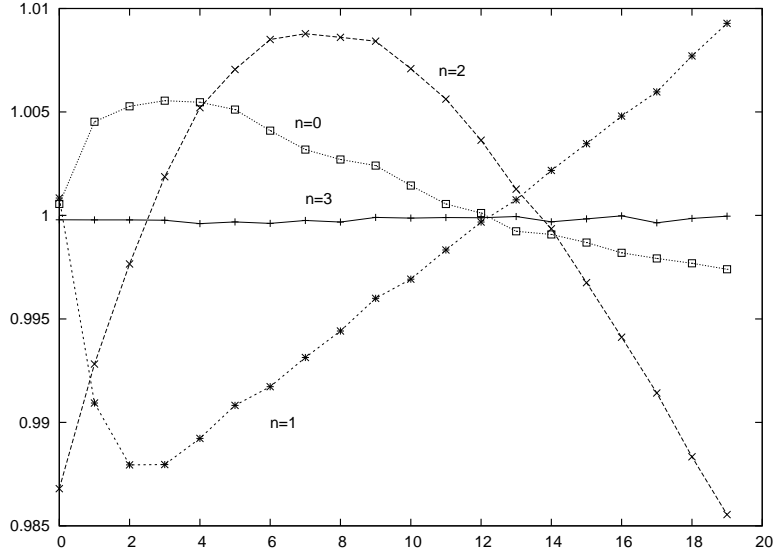}
\caption{\small $n=0,1, 2$ and $3$, $T=0.25, N=20$. $n$ is the samples drawn in vain from the PRNG, $T$ is the period and $N$ is the number of boxes. Dependence on $n$ is related to the correlations among consecutive numbers. \protect\label{fig:graph1}}
\end{center}
\end{figure}

The model we study is simple: There are $N$ boxes numbered $1$ to $N$ in a chain structure, all boxes are equally likely to receive a ball at a random time driven by a Poisson process (Fig.~\ref{fig:time_scheme}). Each time when a ball is received,
a random time increment of $\delta t$ has elapsed, where $\delta t = -\log(r)/N$ and $r$ is a uniformly distributed random number between $0$ and $1$. This then defines a Poisson process in which an event occurs with rate $N$ per unit time. 
With this choice of $\delta t$, on the average $N$ balls are placed in unit time. The box that will receive the ball is chosen randomly as well, such that $i^{\mbox{th}}$ box is chosen if $i-1 <rN < i$  (Fig.~\ref{fig:graph1}). In between these
two steps (choosing the random time and choosing the random box), $n$ samples are chosen from the random number sequence in vain, {\em i.e.}, they are discarded from the used sequence. Furthermore, the simulation is stopped at multiples of 
a time period $T$, so that at each time step when the interval crosses a multiple of $T$. The time variable is set to that time (that is, no balls are placed to any box) and then the simulation continues as described in Fig.~\ref{fig:time_scheme}.
The pseudocode for the model can be seen in the Table 1.\\

\begin{table}[!ht]
\footnotesize
\scalebox{0.95}{
\begin{tabular}{cl||c||c||c||c}
\hline 
\multicolumn{6}{c}{Pseudocode for the process}\tabularnewline
\hline 
 & \multicolumn{5}{l}{1. SET $N$ to lattice size }\tabularnewline
 & \multicolumn{5}{l}{2. SET $period$ to $T$ }\tabularnewline
 & \multicolumn{5}{l}{3. SET $totalTime$ to $0$ }\tabularnewline
 & \multicolumn{5}{l}{4. SET $n$ to number of draws in vain}\tabularnewline
 & \multicolumn{5}{l}{5. SET $a$ to $1$ }\tabularnewline
 & \multicolumn{5}{l}{6. SET occupation of $k^{th}$ site $o(k)$ to $0$ for $ 1 \le k \le N$ }\tabularnewline
 & \multicolumn{5}{l}{7. DRAW a random number $r$}\tabularnewline
 & \multicolumn{5}{l}{8. SET $incrementalTime$ to $-log(r)/N$}\tabularnewline
 & \multicolumn{5}{l}{9. IF $totalTime$ plus $incrementalTime$ is smaller than $a \times T$ THEN }\tabularnewline
 & \multicolumn{5}{l}{10.\rule{1 cm}{0 cm} INCREASE $totalTime \rightarrow totalTime+incrementalTime$}\tabularnewline
 & \multicolumn{5}{l}{11.\rule{1 cm}{0 cm} DRAW a random number $r$}\tabularnewline
 & \multicolumn{5}{l}{12.\rule{1 cm}{0 cm} DRAW $n$ random numbers in vain}\tabularnewline
 & \multicolumn{5}{l}{13.\rule{1 cm}{0 cm} CHOOSE $k^{th}$ box as $k= \lceil rN \rceil$}\tabularnewline
 & \multicolumn{5}{l}{14.\rule{1 cm}{0 cm} INCREMENT $o(k) \rightarrow o(k)+1 $ }\tabularnewline
 & \multicolumn{5}{l}{15. ELSE }\tabularnewline
 & \multicolumn{5}{l}{16.\rule{1 cm}{0 cm} INCREMENT $a \rightarrow a+1$}\tabularnewline
 & \multicolumn{5}{l}{17.\rule{1 cm}{0 cm} SET $totalTime$ as $a \times T$ }\tabularnewline
 & \multicolumn{5}{l}{18. REPEAT $6-17$ steps for $1$ billion times }\tabularnewline
 \hline 
 \label{table:pppseudo}
\end{tabular}
}
\caption{The pseudocode which is neccesary to produce the results in this paper.}

\end{table}

\begin{figure}[!h]
\begin{center}
\includegraphics[scale=0.85]{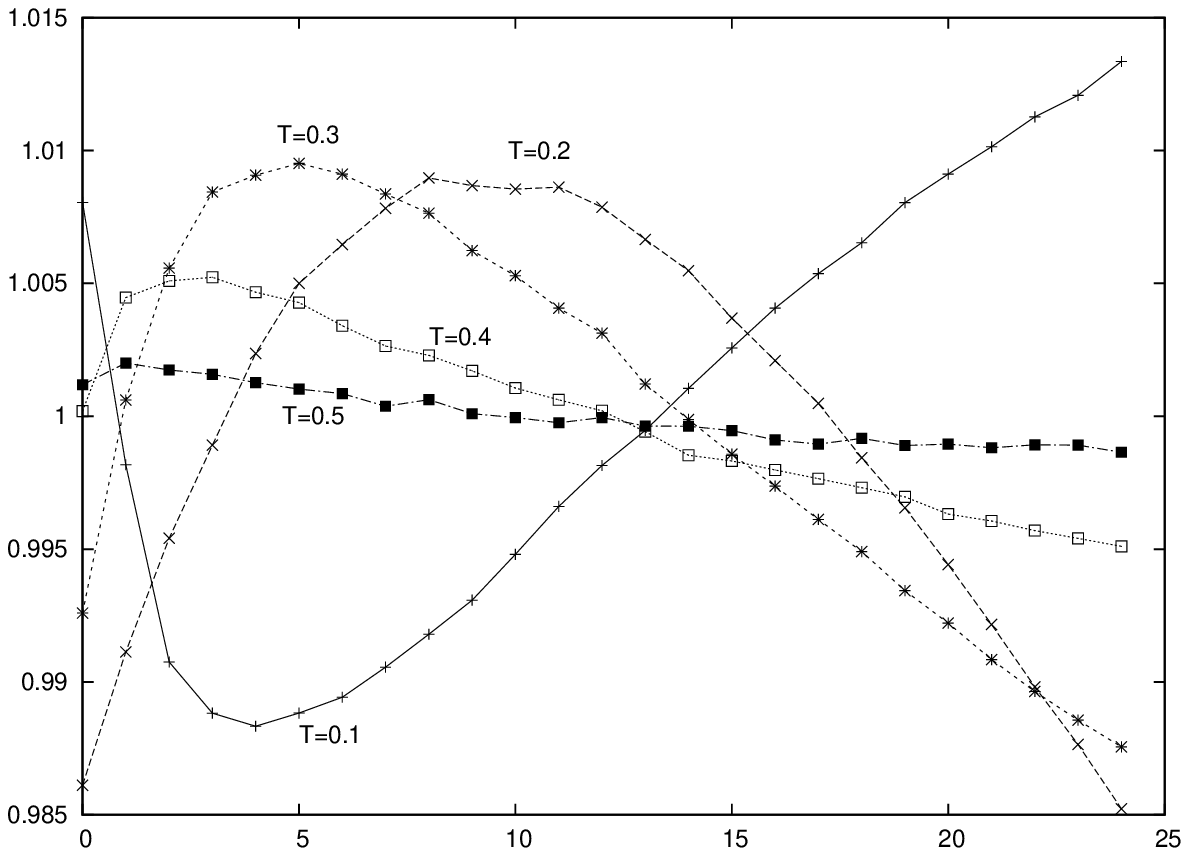}
\caption{\small $n=0, T=0.1, 0.2, 0.3, 0.4$ and $0.5$. For different values of periods we see different shapes of graphs, however for higher periods correlations cease to exist. \protect\label{fig:period1}}
\end{center}
\end{figure}

The process involving the period ($T$) dependence is necessary to observe the correlation. Indeed Fig.~\ref{fig:period1} shows that the process is sensitive on the period. In our research, we were analysing the effect of a periodic perturbation on 
a system, and that is how was the problem discovered. We then tested the process for time increment derived from a simpler form $\delta_t = r/N$, and still found correlations. \\

It is expected that over a large period of time each site will be equally likely to be occupied. However, for $n=0$, boxes in the middle have a tendency to be occupied more than the sites near the both ends (Fig.~\ref{fig:graph1}). Similarly, 
for $n=1$ and $n=2$, correlations are observed. However if $n > 2$,  correlations vanish, in other words there are no preferences left among the boxes. Note that the induced correlations are significant, i.e., they exceed $1\%$ in our example. 
Also note that, simulations are done for $10^9$ trials, which is  well below the period of the PRNG. \\ 
Note that, the abscissa of the graphs in Figures \ref{fig:graph1} and \ref{fig:period1} correspond to the size of the random number used for selecting a box. Different shapes of graphs in these plots correspond to a non-uniform distribution for the
size of the random number. Dependences on $n$ and $T$ are then related to the correlations among the random numbers. 
  
\begin{figure}[!h]
\begin{center}
\includegraphics[scale=0.85]{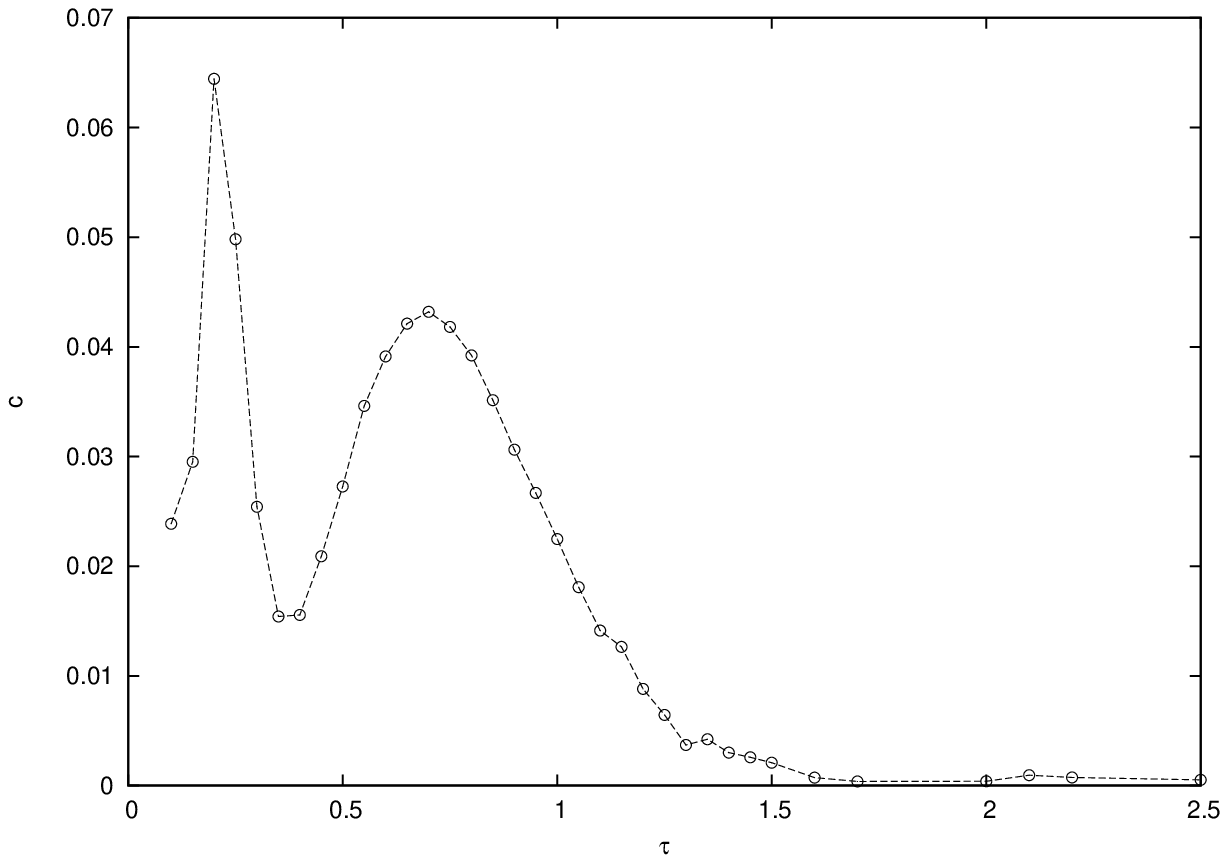}
\caption{\small For different values of period the magnitude of first Fourier coefficient $c$. It represents the size of the systematic error as a function of the period. \protect\label{fig:coeff}}
\end{center}
\end{figure}

Since we wanted a measure of the systematic error in the probability distribution $P$, we looked
at the first coefficient of the Fourier expansion of this distribution: $c=\sum_{j=1}^N{exp(i\pi j/N)(P(j)-c_0)}$ where $c_0$ is the average value of $P$. Figure \ref{fig:coeff} shows magnitude of $c$, which represents the size of the systematic
error as a function of the period.

\pagebreak[4]

\section{Conclusions}
Given the common usage of C as a scientific programming language, we report of correlations in the default random number generator of Glibc. Even though it can pass some tests, PRNG of Glibc shows significant correlations under this very 
simple test. Systematic errors of this magnitude will cause problems in simulations which demand random numbers with low correlations, such as Monte Carlo analysis. Scientific community should beware of accepting default random number 
generators even for well established, commonly used programming languages as C.  We could not identify the precise mechanism which leads to this behavior. A mathematical analysis of this mechanism represents an interesting challenge for 
further analysis.

\section{Acknowledgements}
This work was supported by Turkish Academy of Sciences.

\end{document}